\documentclass{iopjournal}
\usepackage{graphicx} 
 \usepackage{amsmath} 
 \usepackage{enumitem}

\begin{document}

\articletype{Paper} 

\title{Using Consumer Cameras to Observe Scintillation Light from Radiation}

\author{Yuzuka~Sasaki$^1$, Yuuki~Wada$^{2,3}$\orcid{0000-0001-8953-3345} and Kazuo~S.~Tanaka$^{4,5*}$\orcid{0000-0001-6916-9654}}
\affil{$^1$United World College of the Adriatic, Duino, Italy}
\affil{$^2$Faculty of Science, Hokkaido University, Sapporo, Japan}
\affil{$^3$Graduate School of Engineering, The University of Osaka, Suita, Japan}
\affil{$^4$Waseda Research Institute for Science and Engineering, Waseda University, Tokyo, Japan}
\affil{$^5$Accel Kitchen LLC, Sendai, Japan}
\affil{$^*$Author to whom any correspondence should be addressed.}
\email{info@accel-kitchen.com}
\keywords{Radiation measurement, Scintillator,  Gamma-ray, Consumer camera}

\begin{abstract}
For a long time, the cloud chamber was the only educational tool available for measuring radiation. In recent years, simple radiation detectors combining scintillators with silicon photomultipliers have become increasingly common for these purposes. However, students are not able to see the scintillation light, the core process of radiation measurements with scintillators. Therefore, we explored the possibility of detecting scintillation light using two general‑purpose cameras. In addition, we examined how differences in the spatial distribution relate to radiation types and energies. Scintillation light were able to be measured by a general-use camera, and their spatial distribution indicates radiation energy. This method could be utilized as an accessible imaging setup to compare radiation properties in a classroom.

\end{abstract}

\section{Introduction}
Hands-on experiences are essential for intuitively understanding invisible radiation. For a long time, the only educational tool available for directly visualizing radiation was the cloud chamber, in which the passage of radiation leaves visible tracks in an alcohol vapor. In recent years, however, simple radiation detectors combining scintillators with silicon photomultipliers (SiPMs) have become increasingly common in education and outreach activities \cite{Axani2018CosmicWatch}. A scintillator is a material that emits faint flashes of light when struck by radiation, and these weak signals are detected by a highly sensitive light sensor such as an SiPM.

For example, at Accel Kitchen, more than 300 radiation detectors equipped with a scintillator and an SiPM have been distributed to high school students, enabling a wide range of student-led research activities. These include the observation of cosmic rays and environmental radiation \cite{Enomoto2024OnlineSupport}, as well as the construction of Cherenkov detectors by replacing the scintillator with an acrylic block \cite{Tanaka2024RadiationDetectors}. Such initiatives have greatly expanded opportunities for students to engage in radiation-related investigations.

Because scintillation light is extremely faint, it typically requires specialized light-amplifying sensors such as SiPMs or traditional photomultiplier tubes. This has been a major barrier to implementing such experiments in typical school settings. In addition, these devices allow students to detect radiation counts in real time, but the scintillation light itself, the core physical process inside the detector, remains hidden from the student. This presents a missed opportunity: while students can read numbers on a display or hear audible clicks when detecting radiation, they do not experience the fundamental phenomenon that the detector is based on. If students could observe the faint flashes of scintillation light directly, they would gain a deeper conceptual understanding of how scintillator detectors work and why they are widely used in both fundamental research and applications. Meanwhile, the sensitivity of digital cameras, typically equipped with charge-coupled device (CCD) or complementary metal-oxide-semiconductor (CMOS) sensors have greatly improved in recent years. This inspired us to explore whether consumer digital cameras, already commonly available in schools, could be used to capture the faint scintillation light without the need for dedicated photodetectors such as SiPMs.

In this paper, we present a classroom-ready activity where students image scintillation light in a cesium iodide (CsI) scintillator with consumer cameras and analyse long-exposure images to compare one-dimensional intensity profiles of scintillation for different radionuclides. The spatial patterns reveal clear qualitative differences related to radiation type and energy, enabling students to distinguish gamma sources at a glance. By directly visualising scintillation light, students can connect the abstract concept of radiation detection to a tangible observation, strengthening both engagement and conceptual understanding.

\section{Method}
\subsection{Experimental setup}
\begin{figure}[htbp]
 \centering
        \includegraphics[width=\textwidth]{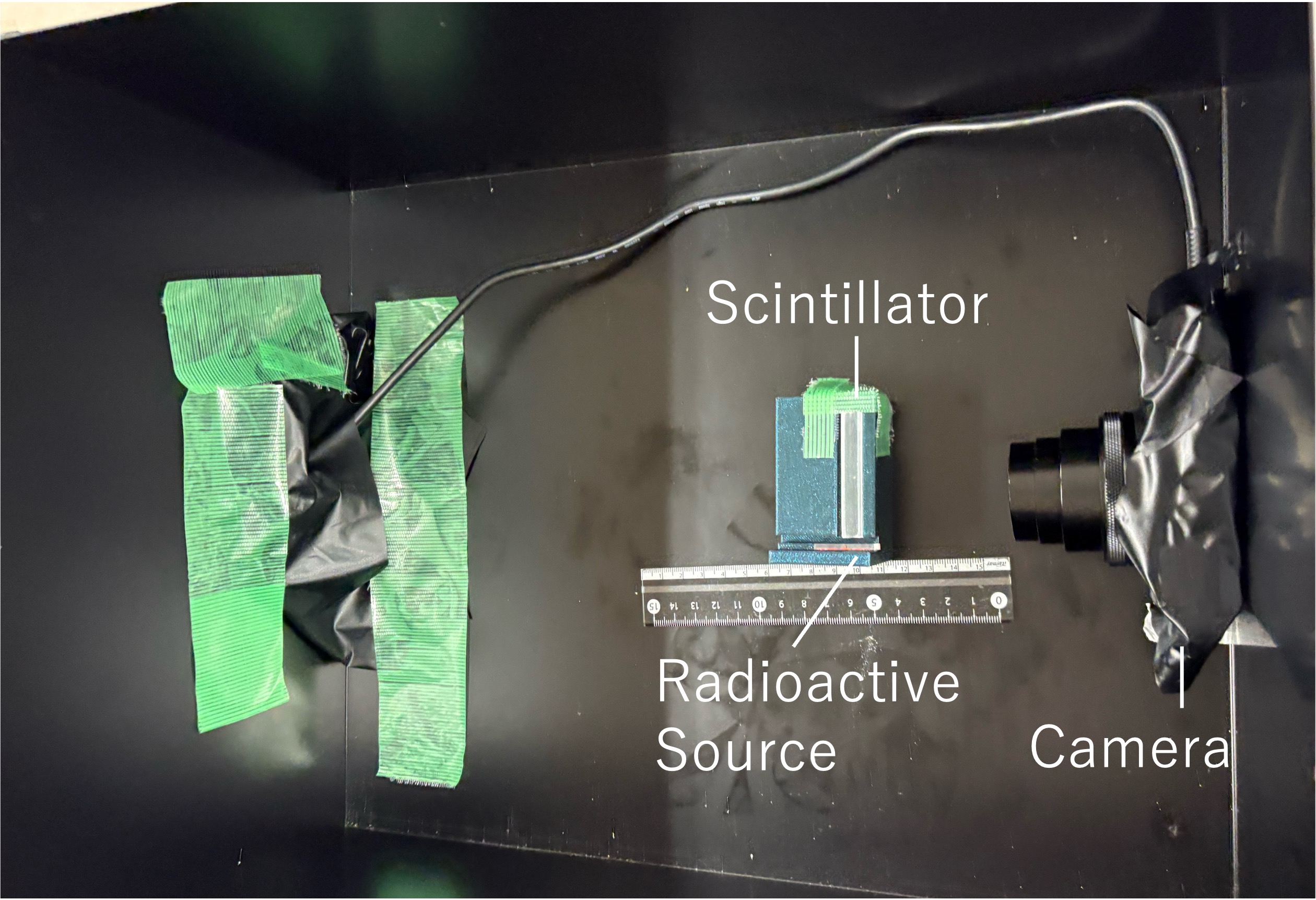}
 \caption{The measurement setup in the light‑tight enclosure. The camera images the CsI scintillator with sources mounted at the left edge. The cooled CCD camera was used in an identical arrangement.}
\label{setup.jpg}
\end{figure}

The experiment was conducted inside a light-tight enclosure containing a CsI scintillator (49$\times$10$\times$6~mm$^3$) and a camera system. The image was obtained from the side with 49$\times$10 mm$^2$.
Three radioactive sources sealed by acrylic resin were used: Am-241, Ba-133, and Cs-137.
The properties of those sources, including activities, are listed in Table~\ref{tab:radioactive_sources}.
Each source was placed on the left side of the scintillator, as shown in Fig.~\ref{setup.jpg}.

Two types of cameras were tested: a digital camera and a cooled CCD camera.
The digital camera (SONY DSC-RX100M7) was operated in bulb mode (continuous exposure mode) with an aperture of F2.8 and an ISO sensitivity of 12800.
For each source, a single exposure of approximately 600~s was acquired and saved as a 16-bit colored TIFF image at 5472$\times$3648~pixels.

The cooled CCD camera (ZWO ASI533MM Pro) was equipped with a lens of the 25~mm focal length and the F1.8 aperture (7artisans, model 25M43B).
It acquired 16-bit grayscale TIFF images at 3008$\times$3008~pixels, also with a single 600-s exposure.
To prevent unwanted reflections of scintillation light, the aluminum plate was wrapped in a light-blocking cloth.

\begin{table}
\caption{Three sources used for the experiment}
\centering
\begin{tabular}{l c c c}
\hline
Nuclide & Am-241 & Ba-133 & Cs-137 \\
\hline
Decay mode & alpha decay & electron capture decay & beta decay \\
Gamma ray energy [keV]\\(emission probability) & 59.54
(35.9\%)
 & 356.0
(62.0\%)
80.99
(32.9\%)
 & 661.6 (85.1\%) \\
Activity [kBq] (on 2025 June)  & 375 & 70.1 & 192 \\
\hline
\end{tabular}
\label{tab:radioactive_sources}
\end{table}

\subsection{Image Preprocessing}
The acquired images were processed in three steps:
\begin{enumerate}[label=(\roman*)] 
    \item \textbf{Cropping:} 
    We retained only the region near the scintillator. 
    
    \item \textbf{Noise reduction:} 
    Median filtering was applied to reduce noise caused by long exposure (OpenCV function cv2.medianBlur, kernel size=3).
    
    \item \textbf{Linearization:} 
    Because stored pixel values in common image file formats do not scale linearly with light intensity, 
    we converted standard RGB (sRGB) values to linear intensity using Eq.~\ref{eq:gamma_correction} from \cite{sRGB}. 
    Since the images used in this study are 16-bit data, channel values were first normalized by 65535. 
    Let $x$ denote a normalized pixel value and $X$ denote the corresponding linear light intensity, 
    which we refer to as luminance.
\end{enumerate}

\begin{equation}\label{eq:gamma_correction}
  X =
  \begin{cases}
    \frac{x}{12.92} \quad (x \leq 0.04045) \\[6pt]
    \left( \frac{x + 0.055}{1.055} \right)^{2.4} \quad (x > 0.04045)
  \end{cases}
\end{equation}

\section{Results}
\begin{figure}[htbp]
 \centering
    \includegraphics[width=\textwidth]{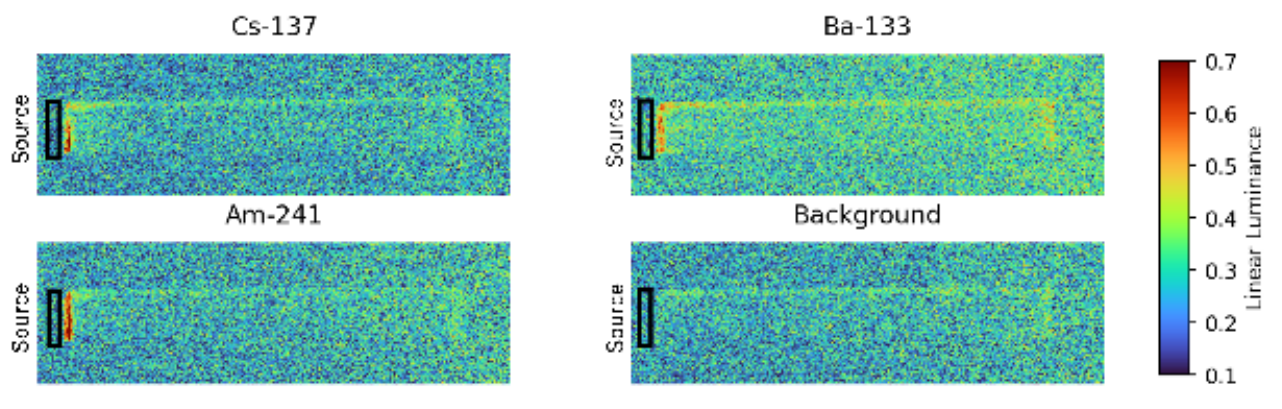}
 \caption{
 Image obtained with a 600-second exposure using the digital camera.
 Luminance was calculated based on Eq.~(1).
 Scintillation light is visible on the left side where the source was placed.
 The region closer to the source is brighter, and the width of the bright region decreases
 in the order of Cs-137, Ba-133, and Am-241, indicating that higher-energy X-rays and gamma rays
 penetrate farther into the scintillator.
 }
 \label{fig:digital_camera_heatmap}
\end{figure}

\begin{figure}[htbp]
 \centering
    \includegraphics[width=\textwidth]{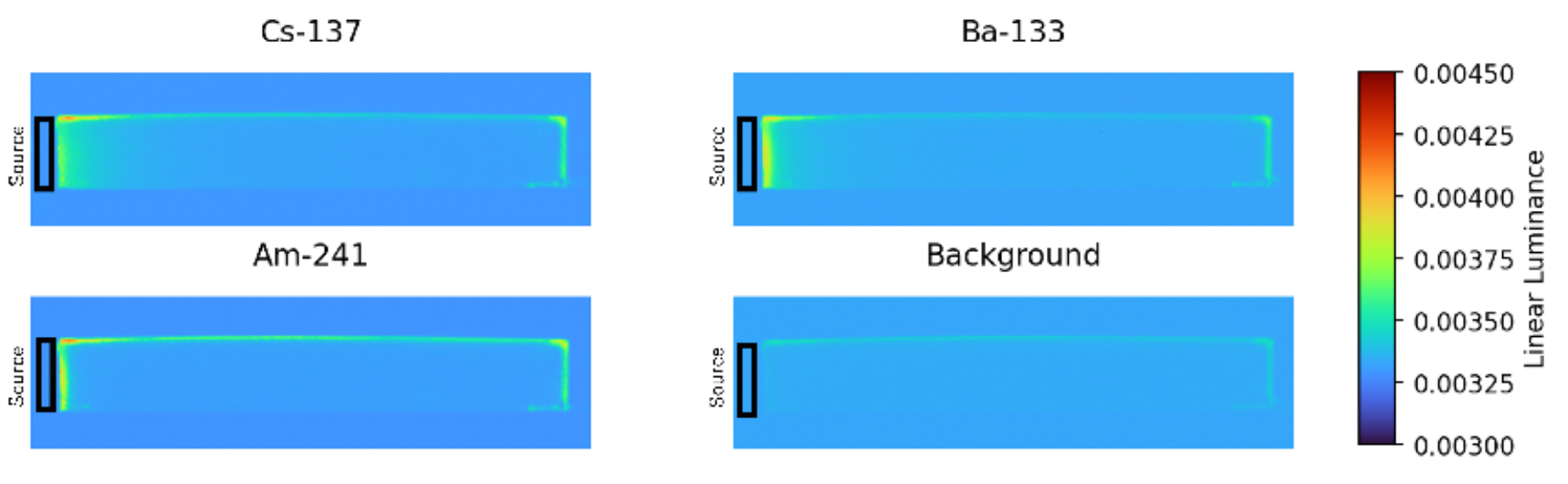}
 \caption{
 Image obtained with a 600-second exposure using the cooled CCD camera.
 Luminance was calculated based on Eq.~(1).
 Scintillation light is more clearly visible than with the digital camera,
 showing a higher signal-to-noise ratio and clearer definition of the bright region near the source.
 }
 \label{fig:ccd_camera_heatmap}
\end{figure}

Figures~\ref{fig:digital_camera_heatmap} and \ref{fig:ccd_camera_heatmap} show images obtained with a 600-second exposure using the digital camera and the cooled CCD camera, with luminance calculated based on Eq. (1). With both cameras, scintillation emission can be observed on the left side where the source was placed. In particular, the cooled CCD camera clearly captured the scintillation light, showing that the region closer to the source was brighter. Moreover, the width of the bright region decreased in the order of Cs-137, Ba-133, and Am-241. This trend reflects the fact that higher-energy X-rays and gamma rays, as listed in Table \ref{tab:radioactive_sources}, penetrate farther into the scintillator.
\par

\begin{figure}[htbp]
 \centering
 \begin{minipage}[b]{0.48\textwidth}
   \centering
   \includegraphics[width=\textwidth]
   {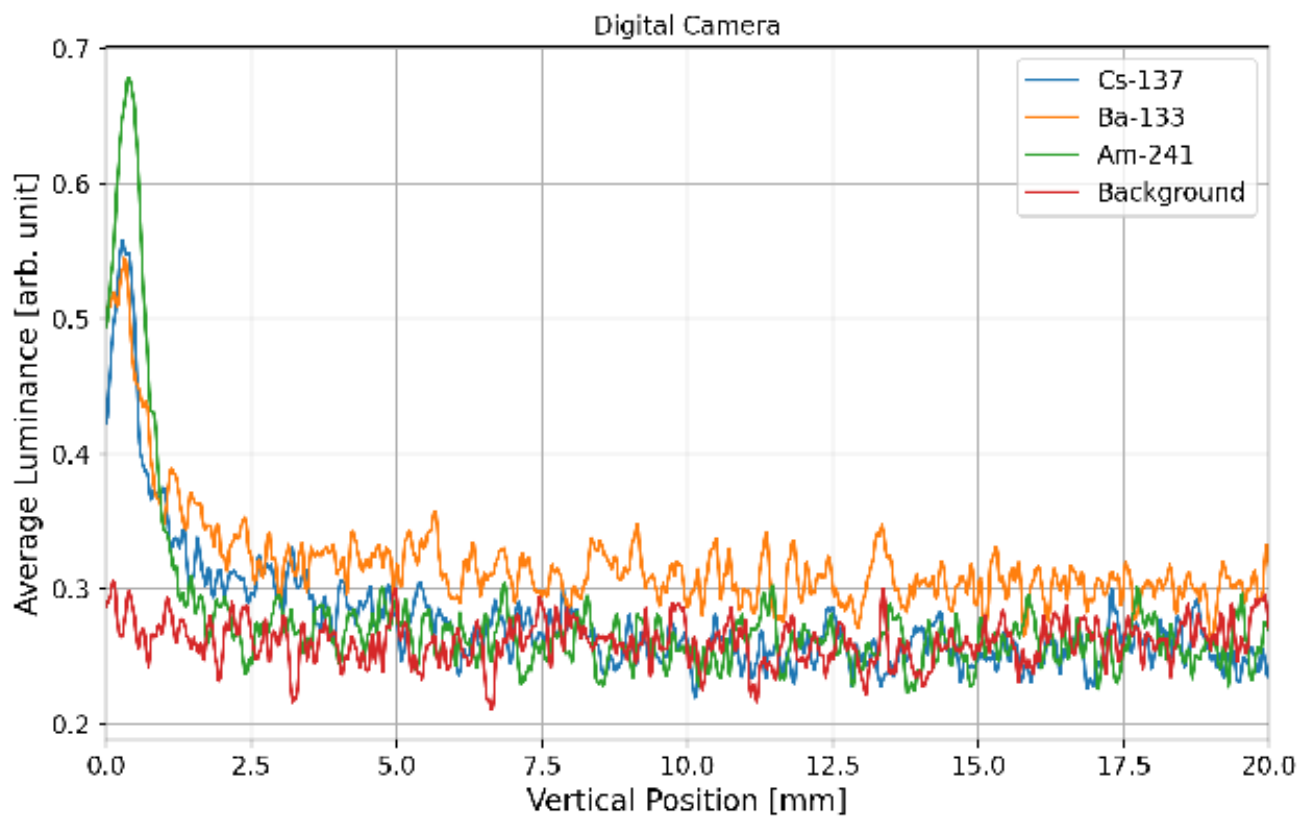}
 \end{minipage}
 \hfill
 \begin{minipage}[b]{0.48\textwidth}
   \centering
   \includegraphics[width=\textwidth]  {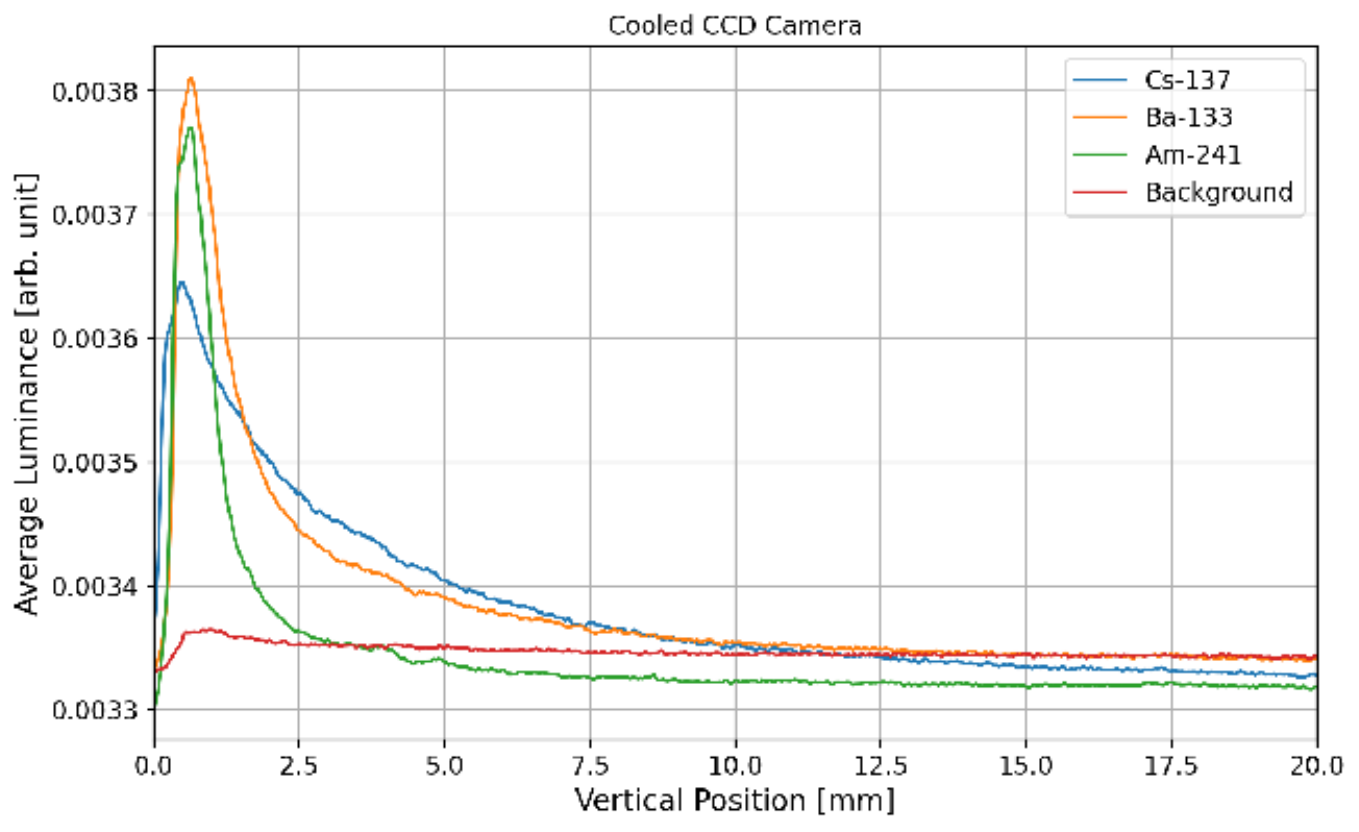}
 \end{minipage}

 \caption{
 One-dimensional intensity profiles plotted as a function of distance, with the source-facing edge defined as 0~mm.
 The cooled CCD camera shows a lower dark current and a higher signal-to-noise ratio compared to the consumer digital camera.
 The profiles from the cooled CCD clearly demonstrate the spread of scintillation light distribution corresponding to the energy of each radiation source.
 }
 \label{line_graph_without_aluminum_plate.eps}
\end{figure}

To quantitatively investigate the distribution of the scintillation light, the one-dimensional intensity profiles as a function of distance, with the source-facing edge set to 0~mm, are shown in Fig. \ref{line_graph_without_aluminum_plate.eps}. Compared to the digital camera, the cooled CCD camera suppressed dark current and achieved a higher signal-to-noise ratio (S/N). In addition, the cooled CCD camera clearly revealed differences in the spread of the scintillation light distribution depending on the energy of radiation from each source. Cs-137, emitting gamma rays with an energy of 661.6~keV, has a wider scintillation light distribution compared to Am-241, which emits gamma rays with an energy of 59.54~keV. This meets the expectation as gamma rays with stronger energy are likely to travel longer distances before interacting. 
\par

\begin{figure}[htbp]
 \centering
        \includegraphics[width=0.8\textwidth]{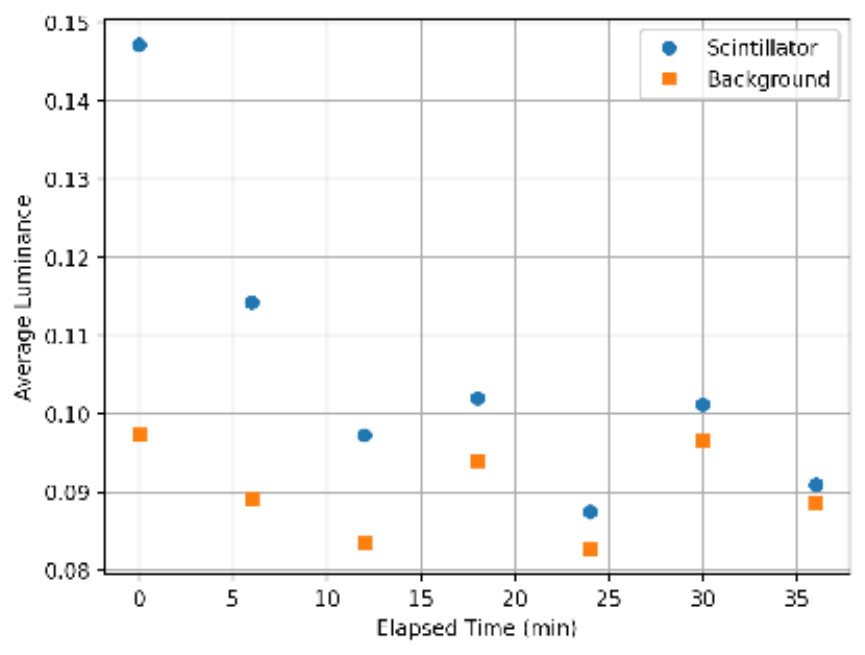}
 \caption{Temporal decay of the average luminance in the scintillator region versus elsewhere in the frame. 300 s exposures followed by 300 s intervals were performed seven times. The difference in average luminance attenuated over time.
}
\label{line_graph_afterglow.eps}
\end{figure}

To assess residual afterglow in the scintillator, the field was kept light‑tight, and successive 300‑s exposures were taken at 300‑s intervals. The average luminance over the scintillator region attenuated towards the background level over time (Fig. \ref{line_graph_afterglow.eps}).

\section{Discussion}
The observed attenuation of scintillation light luminance and its energy dependence are consistent with expectations for gamma-ray attenuation in CsI scintillator. Gamma rays with higher energy (e.g., 661.6 keV from Cs‑137) penetrate materials easily, leading to a wider bright region and a less sharply shaped peak than gamma rays with lower energy (e.g., 59.54 keV from Am-241). 

\begin{figure}[htbp]
 \centering
        \includegraphics[width=0.8\textwidth]{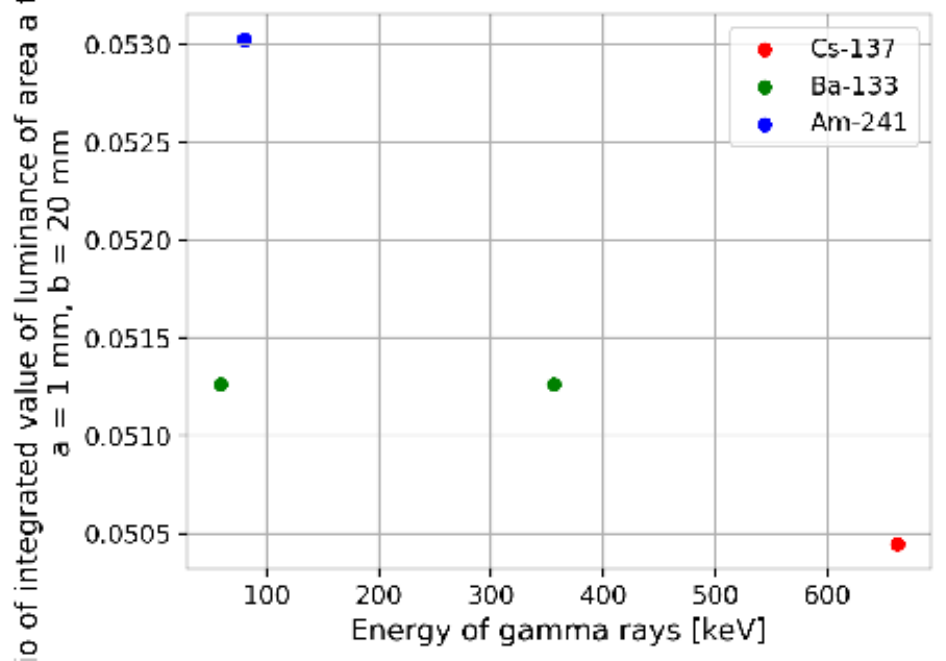}
 \caption{The ratio of the peak area of 0 to1~mm to the peak area from 0 to 20~mm takes the largest value with Am-241 and the smallest value with Cs-137. This indicates that gamma rays with weaker energy have a sharper peak, which is consistent with the expectations that weaker gamma rays attenuate faster in substances.
}
\label{ratio_peak_area.eps}
\end{figure}

\begin{figure}[htbp]
 \centering
        \includegraphics[width=0.8\textwidth]{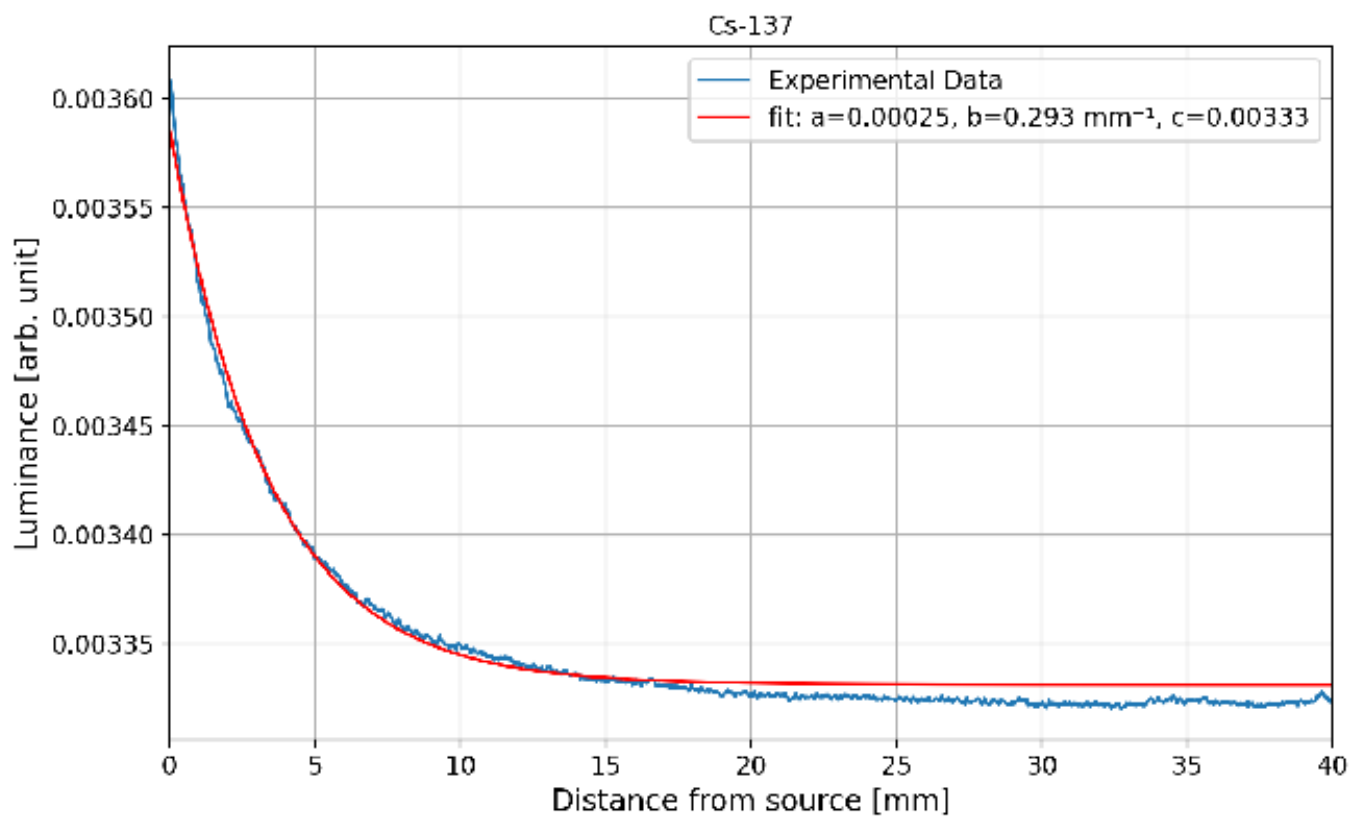}
         \includegraphics[width=0.8\textwidth]{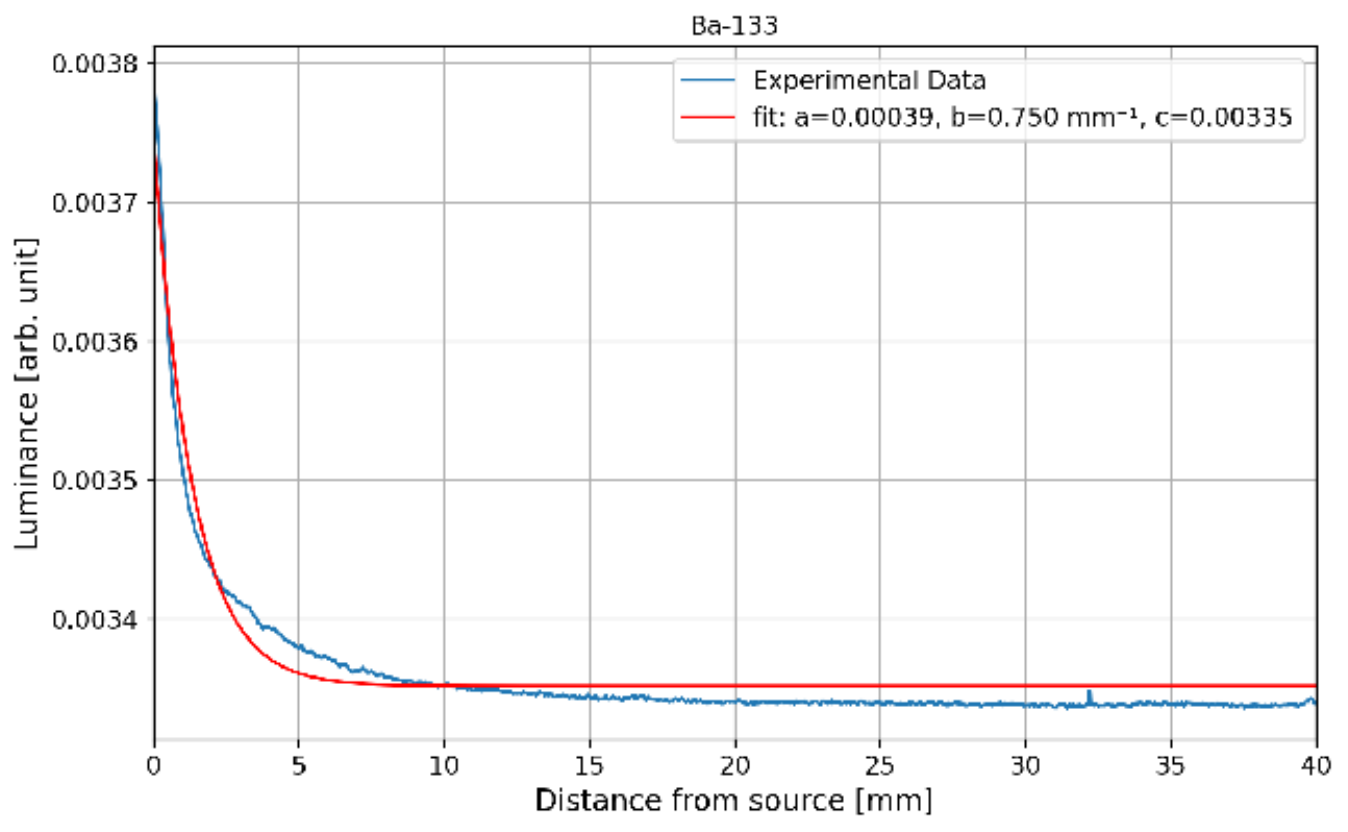}
         \includegraphics[width=0.8\textwidth]{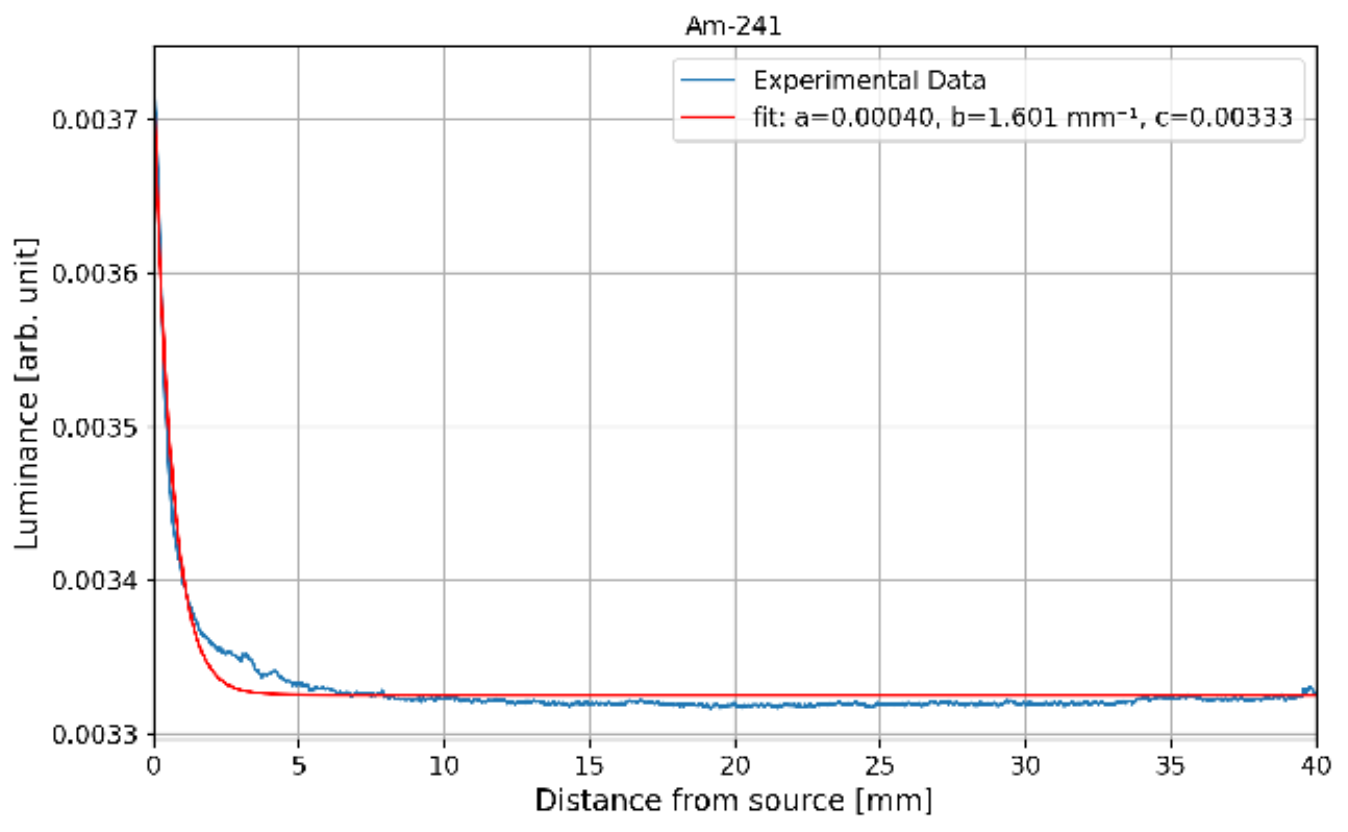}
 \caption{One-dimensional intensity profile was fitted with the function $f(x) = a*exp(-bx)+c$. The value of b takes the largest value with Am-241 and the smallest value with Cs-137. This shows that gamma rays with stronger energy attenuate gradually compared to those with smaller energy.
}
\label{fitting.eps}
\end{figure}

The ratio of the peak area from 0 to 1 mm to the peak area from 0 to 20 mm for each source is compared in Fig. \ref{ratio_peak_area.eps}. The ratio decreases as the energy of the gamma rays decreases. This indicates that gamma rays with weaker energy have a sharper peak, which is consistent with the expectations that weaker gamma rays attenuates faster in substances.
As the intensity of gamma rays decays exponentially, fitting of the one-dimensional intensity profile with the function
\begin{equation}\label{eq:fitting equation}
  f(x) = ae^{-bx} +c
\end{equation}
was conducted. The result is shown in Fig. \ref{fitting.eps}. The values of b decrease as the energies of the gamma rays increase. This matches the expectation that gamma rays with stronger energy attenuate gradually compared to those with smaller energy. 
These trends indicate that spatial scintillation patterns measured using consumer digital cameras can serve as a simple way to determine radiation energy and type.

\section{Conclusion}
Using a general-purpose digital camera and a cooled CCD camera, we recorded scintillation light from a Cesium Iodide scintillator induced by three different sources. The spatial distribution of luminance induced by different radionuclides revealed energy-dependent patterns. These findings suggest that this method could be an accessible imaging setup for students to compare radiation properties.

\section{Safety and Risk Assessment}
The sources used in this activity are sealed sources listed in Table 1, each containing radioactivity below the legal threshold and therefore not subject to regulation under the Act on the Regulation of Radioisotopes, etc. The sources were kept in a locked, clearly labelled storage container and handled only under the supervision of a trained instructor.
This demonstrates that even elementary, junior high, and high schools can safely use such sources to produce sufficient scintillation light observable with an ordinary consumer camera.
\section{Acknowledgements}
This work was supported by the SEEDS program of The University of Osaka, and The Mitsubishi Memorial Foundation for Educational Excellence.

\bibliographystyle{unsrt}
\bibliography{reference}

\end{document}